# Thermoelectric imaging of structural disorder in epitaxial graphene

Sanghee Cho[1]†, Stephen Dongmin Kang[1]†, Wondong Kim[1], Eui-Sup Lee[2], Sung-Jae Woo[2], Ki-Jeong Kong[3], Ilyou Kim[4], Hyeong-Do Kim[4]‡, Tong Zhang[5]‡, Joseph A. Stroscio[5], Yong-Hyun Kim[2]\*, Ho-Ki Lyeo[1]\*

[1]Korea Research Institute of Standards and Science, Daejeon 305-340, Republic of Korea

[2]Graduate School of Nanoscience and Technology (WCU), Korea Advanced Institute of Science and Technology, Daejeon 305-701, Republic of Korea

[3]Korea Research Institute of Chemical Technology, Daejeon 305-600, Republic of Korea

[4]Pohang Accelerator Laboratory, Pohang University of Science and Technology, Pohang 790-784, Republic of Korea

[5]Center for Nanoscale Science and Technology, National Institute of Standards and Technology, Gaithersburg, MD 20899, USA

†These authors contributed equally to this work.

‡Present addresses: Center for Functional Interfaces of Correlated Electron Systems, Institute for Basic Science, Seoul National University, Seoul 151-747, Republic of Korea (H.-D.K.); Department of Physics, Fudan University, Shanghai 200433, China (T.Z.).

\*e-mail: hklyeo@kriss.re.kr; yong.hyun.kim@kaist.ac.kr

**Heat is a familiar form of energy transported from a hot side to a colder side of an object, but not a notion associated with microscopic measurements of electronic properties. A temperature difference within a material causes charge carriers, electrons or holes, to diffuse along the temperature gradient inducing a thermoelectric voltage. Here we show that local thermoelectric measurements can yield high sensitivity imaging of structural disorder on the atomic and nanometre scales. The thermopower measurement acts to amplify the variations in the local density of states at the Fermi-level, giving high differential contrast in thermoelectric signals. Using this imaging technique, we uncovered point defects in the first layer of epitaxial graphene, which generate soliton-like domain wall line patterns separating regions of the different interlayer stacking of the second graphene layer.**

Materials synthesis involves a variety of processes that generate stress in the solid. Materials respond to stress in different ways, creating a diversity of structural imperfections and strain in the lattice[1], which have their origins at the atomic level. Detecting these subtle features with high sensitivity and resolution, and their influence[2] on electronic, thermal, and mechanical properties, remains challenging. Scanning probe methods, particularly tunnelling microscopy[3], have been useful to measure such properties locally. However, standard measurements are not well suited to finding the small changes in lattice parameters that accompany strain and disorder, and large-scale imaging of the associated electronic states is prohibitive. In this report, we show how microscopic thermopower measurements can image structural changes in a material, which works by detecting modifications in local electronic properties.

The high temperature (>1100 °C) processing of epitaxial graphene on SiC, which offers a unique avenue to large-scale production of graphene electronics[4,5], leads to abundant structural imperfections[6] and strain[7]. Graphene is distinctive in possessing a negative thermal expansion coefficient, and thus a compressive strain, up to 0.8 %, is generated when grown on SiC (ref. 8). Although the concrete response behaviour of graphene to this strain energy remains elusive, partial relaxation[7], mediated through defects[8,9], has been recognized to leave a non-uniform distribution of residual strain[7]. Understanding this disorder and strain is essential to tailoring new graphene electronics because of their influence on electronic properties[6,7,10]. Strain in graphene can also be used to generate pseudo-magnetic fields[11,12].

Disorder and strain can have large and spatially localized electronic signatures[10], even for subtle changes in atomic positions. We detect these signatures with an enhanced sensitivity by exploiting thermopower, an electronic transport property that reveals distortions in the electronic structure. Thermopower reflects the asymmetry[13,14] in the density of states (DOS) with respect to the Fermi-energy, $E_F$ (see Supplementary Information). If thermopower works as a probe of local electronic properties[15-17], structural defects and local deformations, especially those that distort the local DOS near $E_F$, would appear with a contrast in thermopower imaging. Indeed, by imaging thermopower, we discovered a defect-mediated dimensional evolution of strain-response patterns in epitaxial graphene with increasing thickness. Furthermore, atomic-scale local DOS variations were imaged by using this method. These results show the powerful nature of imaging thermoelectric signals.

To develop thermoelectric scanning microscopy, we used a modified ultra-high vacuum atomic force microscope (UHV-AFM) with a gold-coated conducting probe that scans the sample in contact mode while maintaining a temperature difference between the tip and the sample (Fig. 1a). A localized temperature gradient is induced in the vicinity of the tip, yielding a thermoelectric voltage proportional to the local thermopower of the contact region. The resolution of the image depends on the tip-sample contact, and is not necessarily restricted by the areal size of the localized temperature gradient[16,18]. In addition to the thermoelectric voltage, we record the $z$-displacement of the cantilever to simultaneously image both the thermopower and topology of the sample surface, which is particularly advantageous in distinguishing topographic and thermoelectric DOS-induced features. This capability of simultaneous imaging is a significant advancement of the previous prototype based on a scanning tunnelling microscope (STM), which was used to profile thermopower across GaAs p-n junctions[16].

The primary investigation on graphene was conducted by "normal scanning", where we collect transient thermoelectric signals at a scan time comparable to that of conventional AFMs. Follow ups on more quantitative measurements were later performed through single-point thermoelectric spectroscopic measurements (Fig. S4). The thermopower image of epitaxial graphene acquired by normal scanning (Fig. 1c) clearly shows the coexistence of three contrasted regions with different thermopower and resident features, which were identified as monolayer, bilayer, and trilayer graphene (MLG, BLG, and TLG). The enormous detail in the thermopower image is absent in the simultaneously obtained topographic image (Fig. 1c), which only displays the terraced structure of the SiC substrate. The comparison between Fig. 1b,c demonstrates how the topographic information is eliminated in the thermopower image. The complex patterns found in each graphene layer, for example, the dark lines found in BLG, are not terrace steps but strain-related features existing within the graphene layer, as discussed below. The thermopower images of MLG and BLG both produce a negative thermoelectric voltage signal, which is a result of the electron doping induced by the underlying substrate (Fig. S1), as shown by angle-resolved photoemission spectroscopy (ARPES) measurements in Fig. S5.

Figure 2b,d,f shows close-up scans of the complex features in the thermopower images, where the patterns evolve from spots in MLG into line patterns in BLG. Despite their strikingly explicit contrast in the thermopower images, these patterns are either subtle or unobservable in the topographic images (Fig. 2a,c,e,g). The dark line patterns discovered in BLG (Fig. 2d,f) are one of the most conspicuous features with a magnitude of the local thermopower in these line patterns larger than that of the BLG background signal. Features associated with the dark lines are local spots at the joining points of the line feature (Fig. 2d,f). Most of the time, dark spots were observed in the thermopower image at the joining points of lines with different orientations. Many of these spots were also detected in the topographic image as buckled sites in the graphene with a protrusion up to 5 Å (Fig. 2e). In MLG, only the isolated spot features were found, without any extended line features (Fig. 2b). On the other hand, the characteristic feature found in TLG was a patchwork of planar domains (Figs 2g,h and S2), in addition to the spots and lines.

The variation of the patterns, found to be dependent on the number of graphene layers, is reminiscent of the stress relaxation patterns observed in metal epitaxy studies[19,20]. It is well known from these studies that the dimensionality of the stress relaxation patterns in metal epitaxy increases as the overlayer grows. For instance, Cu overlayers on Ru(0001) substrates exhibit a pseudomorphic structure for monolayers, and then are followed by one- and two-dimensional relaxation structures with increasing layer-numbers[19]. The evolution of the surface structure indicates that the phenomenon is driven by the free energy of the system, mainly the strain energy and the inter-layer cohesive energy[19,20]. In this regard, the interaction energy between the topmost graphene layer and its underlying layer, which tends to decrease with increasing layer thickness (see Supplementary Information), and the compressive strain energy in the graphene layer are considered responsible for the formation of the patterns found in the thermopower maps.

We reproduced the tendency in pattern evolution by simulating the response patterns of graphene under compressive strain as a function of the interlayer cohesive energy (which is dependent on the thickness of graphene films) with the Metropolis Monte Carlo algorithm (See Supplementary Information). Figure 2i-l illustrates how a (40 x 40) unit-cell graphene system containing two defect sites, which have weaker interlayer cohesive energy than other sites, accommodates a given uniform compressive strain into a non-uniform distribution. As the cohesive energy with the underlayer decreases with increasing graphene thickness, the relaxation patterns extend from spots to lines connecting the spots and to domains. In other words, graphene prefers to deform locally near the defect sites when it is strongly bound, as in the case of MLG on top of the buffer layer. Localized deformation is indeed observed in some closely examined topographic images of our samples, as shown through the buckled sites (Fig. 2e). These results indicate that the strain energy is locally concentrated at defect sites as a way to lower the total strain energy of the system. Generally, the observed response will be strain patterns following the overall trends in Fig. 2i-l. It is also likely that the line patterns may contain subtle in-plane deformations to lower the total strain energy. In particular, we postulate that the line patterns in BLG may represent strained boundary walls between different AB (Bernal) stacking domains (Fig. 2m), as discussed below. The domains found in TLG are attributed to different stacking sequences[21,22], ABA (Bernal) and ABC (rhombohedral) stacking (See Supplementary Information). By contrast, a less localized response for cases of weakly bound graphene, *e.g.* exfoliated graphene, could lead to rippled textures[23].

We now discuss the origin of the contrast mechanism in thermopower imaging and how defects and lattice strain come to appear in thermoelectric measurements. The main contribution to the local contrast in the thermopower image originates from a change[14] in the local DOS near $E_F$. Additional DOS located above (below) $E_F$—when positioned within the range for the thermal broadening of the Fermi-Dirac distribution (width of df/dE~0.1 eV at 300 K)—produces a negative (positive) contribution to thermopower (Fig. 1a). As structural defects in graphene could produce localized states near the Dirac point $E_d$ (ref.10), these defects would appear in thermopower images depending on their energy position relative to $E_F$. For example, $E_F$ values of MLG and BLG are 0.45 eV and 0.3 eV above $E_d$, respectively, in our samples (Fig. S5) and, therefore, dislocation cores with localized states at 0.35 eV above $E_d$ would appear darker than the background signal in BLG, but brighter in MLG — because these states would be above $E_F$ in BLG (negative shift in thermopower), but below $E_F$ in MLG (positive shift in thermopower).

The thermopower is negatively enhanced by more than a factor of two in the line patterns in BLG (Figs. 1c and 2d,f), which indicates the presence of additional local DOS above $E_F$. This additional DOS results from the strained structure of graphene strips between different stacking domains (Fig. 2m) that form the loops of the line patterns. A domain wall can form if the top layer of graphene is shifted, among six equivalent symmetry directions, by one nearest neighbour distance (1.4 Å) relative to the top layer in a neighbouring domain. The resulting soliton-like domain wall will be a strained strip accommodating the different Bernal stacking sequences, namely AB and BA, in each neighbouring domain, and thus will contain transitional stacking sequences[24] (Fig. 2m). The atomic positions in the wall vary only of the order of 1 % (1.4 Å compression in ≈10 nm widths in our measurements), and the variation could become even smaller as the walls further relax by becoming pinned to the defect sites. Such variation would be hardly observable in topographic images as is the case in Figs. 1 and 2. A recent transmission electron microscopy study has shown similar soliton domain walls in BLG[25].

These one-dimensional domain walls can quantize the electron wavefunctions because the strain and local structural changes form energy barriers; the confinement in the wall of width $L$ leads to a quantization in the wave number $k$, i.e. $kL = n\pi$, where $n$ is a positive integer[26]. The additional DOS that contributes to the large change in thermopower in the line patterns could be due to such quantum confined states in the domain-wall boundary. This would be a visual example of enhanced thermopower due to the quantum confinement in a one-dimensional nanostructure, an idea fostered by Hicks and Dresselhaus[27].

The potential of thermoelectric imaging persists at the atomic scale. Shown in Fig. 3 is an atomically resolved image of BLG obtained near room temperature from both topographic (Fig. 3a) and thermoelectric voltage (Fig. 3b) measurements. The AFM image (Fig. 3a) shows atomic corrugation predominantly in one direction, with a smaller amplitude along the corrugated rows, whereas the thermoelectric image in Fig. 3b displays atomically resolved carbon hexagons in addition to the contrast related to the $6\sqrt{3} \times 6\sqrt{3}R30°$ reconstruction of the underlying SiC lattice, reminiscent of STM images of this surface[28,29] (See Fig. S10). These atomic variations were observable only when there was a finite temperature difference between the probe and the sample. This extremely localized probing is dependent on a sharp probe contact[18,30] and the coherence in thermoelectric transport[15] (further discussions will be given elsewhere). We note that the atomic resolution imaging by STM also relies on the local DOS, but with a different manner of dependency.

The atomic-scale imaging allows us to investigate and identify the structural disorder that leads to the line patterns discovered by large-scale thermopower imaging. This procedure is demonstrated in Fig. 4 where spots and line patterns are found at the nanoscale (Fig. 4a,b) and a local defect is found in the close-up image at the joint intersecting line patterns (Fig. 4c,d). The AFM image (Fig. 4c) shows that the number of (zigzag) atomic rows changes across the defect, which indicates the presence of a heptagon-pentagon pair, i.e. a dislocation core[10,31]. This identification of a naturally existing dislocation core supports the suggested model of a soliton domain wall described in Fig. 2m. The simultaneously obtained thermoelectric image (Fig. 4d) exhibits a complex interference pattern originating from electron scattering. The Fourier transform of this scattering image (inset of Fig.4d) suggests intervalley scattering of the electrons with a wave vector approximately equal to the Fermi wave vector[28]. This electronic signature once again indicates that thermopower imaging reflects information of the local DOS.

More quantitative information can be extracted by stationary spectroscopic measurements, where the tip stays at a position for a sufficient amount of time to establish a steady-state signal. As shown in Fig. 5, the thermoelectric voltage measured with this mode exhibited a linear relation with the applied temperature difference $\Delta T$. This linearity for small $\Delta T$ agrees with the thermoelectric nature expected from graphene, although the magnitude is smaller than expected due to a cancelling contribution from the gold probe tip, which takes up a part of $\Delta T$. The dependence on $\Delta T$ is also seen in imaging measurements (Fig. S8).

In contrast to the theoretical expectations[32], quantitative measurements found the thermopower of MLG to be about three times larger in magnitude than that of BLG (Fig. 5). For free standing or exfoliated graphene, the thermopower of BLG is expected to be slightly larger in magnitude than that of MLG (ref. 32). The smaller value of $E_F - E_d$ in BLG would have even enhanced this difference[32,33]. An explanation for the larger thermopower in MLG is an enhanced scattering caused by the $6\sqrt{3} \times 6\sqrt{3}R30°$ reconstruction with respect to the SiC(0001) surface (see Fig. S6).

The striking sensitivity and contrast of the strain-related patterns in epitaxial graphene demonstrates the nature of thermoelectric measurements that amplify the variations in local DOS as if a physically operated differential filter for the electronic states near $E_F$. Combined with the high throughput of our method, with which areas as large as ~2 μm (Fig. S7) can be scanned at the speed of conventional AFM, we can readily detect strain and defect variations within a material and quantify their areal density. For instance, Fig. 1c yields a defect site density on the order of $10^{10}$ cm$^{-2}$, a property challenging to determine with conventional methods.

Many future opportunities are expected with local measurements of thermopower. By controlling the Fermi level of the sample through gating, measurements would be able to reveal information from a larger range of the energy bands. Furthermore, the Fermi-energy-sensitive nature enabled by a small temperature difference may be exploited in other spectroscopic measurements aimed at investigating the electronic structure, inspiring future novel measurement schemes.

## Methods

**Sample growth** The epitaxial graphene samples were prepared by the thermal decomposition method with 6H-SiC(0001) (Figs 2a-d, g, h, 3 and 4) or 4H-SiC(0001) (Figs 1, 2e,f) substrates. The substrates were pre-etched with a gas mixture of $H_2$ (10 %) and Ar (90 %) at $\approx$1500 °C. After transferring the substrates into an UHV chamber, the substrates were Joule-heated up to $\approx$1120 °C to grow epitaxial graphene layers. The pressure during heating was maintained at $<3 \times 10^{-8}$ Pa. The growth was monitored by low-energy-electron diffraction and the thickness was confirmed by ARPES measurements[34]. The resistivity of the substrates was 0.115 $\Omega\cdot$cm (4H-SiC) and 0.082 $\Omega\cdot$cm (6H-SiC). For the samples shown in Fig. 2e-f and S9, the layers were grown in an Ar atmosphere instead of UHV. For the sample in Fig. S9, the layers were grown in a furnace at $\approx$1600 °C. Before the scanning measurements, all samples were annealed at $\approx$700 °C overnight in the measurement chamber for surface cleaning purposes.

**Scanning probe measurements** The scanning probe instrument was operated in a contact mode with a conducting gold-coated (or diamond-coated) tip controlled by a normal-force feedback circuit. The chamber was maintained at a vacuum level of $<10^{-8}$ Pa, and all the measurements were conducted at a set point of <0.5 nN. The thermoelectric voltage between the tip and the grounded sample surface was measured through an independent circuit with a high-impedance electrometer. The sample was heated with a heating stage and the temperatures of the heating stage, sample and reservoir were monitored independently.

**Acknowledgements**

We thank Professor Young Kuk for helpful comments. This work was supported by the Converging Research Center Program of MEST (2012K001307) and the MEST-US Air Force Cooperation Program of NRF/MEST (2010-00303). The work at KAIST was supported by the WCU (R31-2008-000-10071-0) and NRF (2012-046191) programs of MEST.

**Author Contributions**

S.C., S.D.K. and H.-K.L. performed the experiments, collected and analysed the data, and prepared the manuscript. W.K., I.K., T.Z. and J.A.S. prepared and characterized the samples. E.-S.L., S.-J.W., K.-J.K. and Y.-H.K. conducted the calculations. H.-D.K. and I.K. carried out the ARPES measurements. H.-K.L. and Y.-H.K. administrated the research. H.-K.L., Y.-H.K. and J.A.S. edited the manuscript.

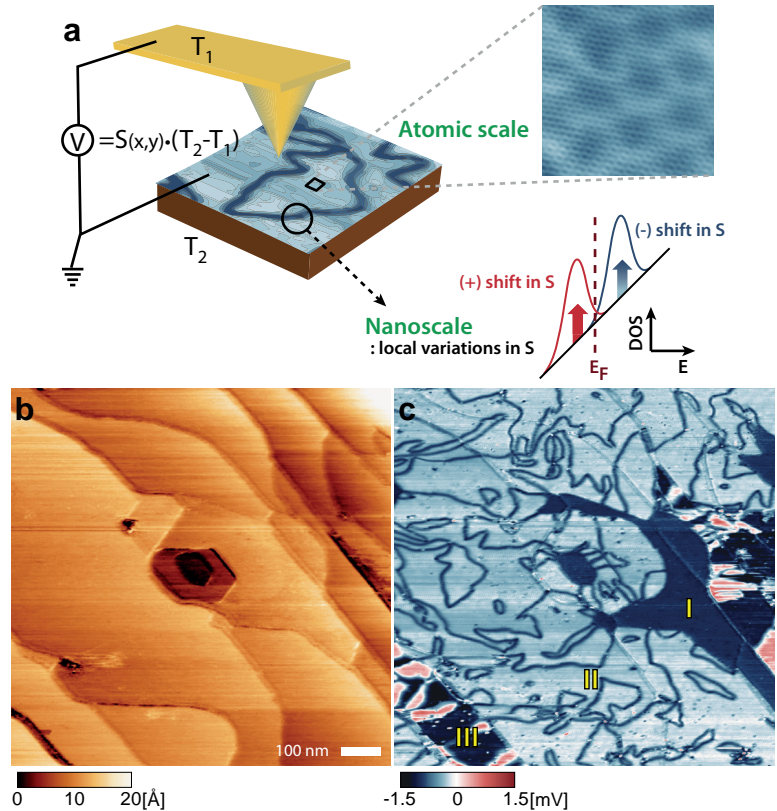

**Figure 1 | Thermoelectric imaging and its application to epitaxial graphene**. **a,** Schematic of the thermoelectric imaging method. A small temperature difference between the gold-coated tip and the sample is maintained by keeping the entire sample slightly heated, inducing a thermoelectric voltage which we record with a voltmeter. When the probe tip encounters any local features at the nanoscale that accompany a local change in the DOS in the proximity of the Fermi level, *e.g.* additional DOS from defects, a local variation in thermopower is detected, resulting in contrast in the image. The thermopower shifts negatively or positively if the increase in DOS is above or below the Fermi level, respectively. Simultaneously, the deflection of the cantilever is recorded to acquire the topological information from the same area of the sample. By adjusting the probe conditions, the imaging method can also be extended down to the atomic scale (see Figs 3 and 4). **b,c** AFM topographic height image **(b)** and thermopower image **(c)** simultaneously obtained from epitaxial graphene. The three distinct regions indexed as I, II, and III correspond to MLG, BLG and TLG, respectively. The temperature difference between the cantilever and the sample was approximately 30 K. Although the image is dominated by the local changes in thermopower, variations in the local temperature profile can modify the observed contrast.

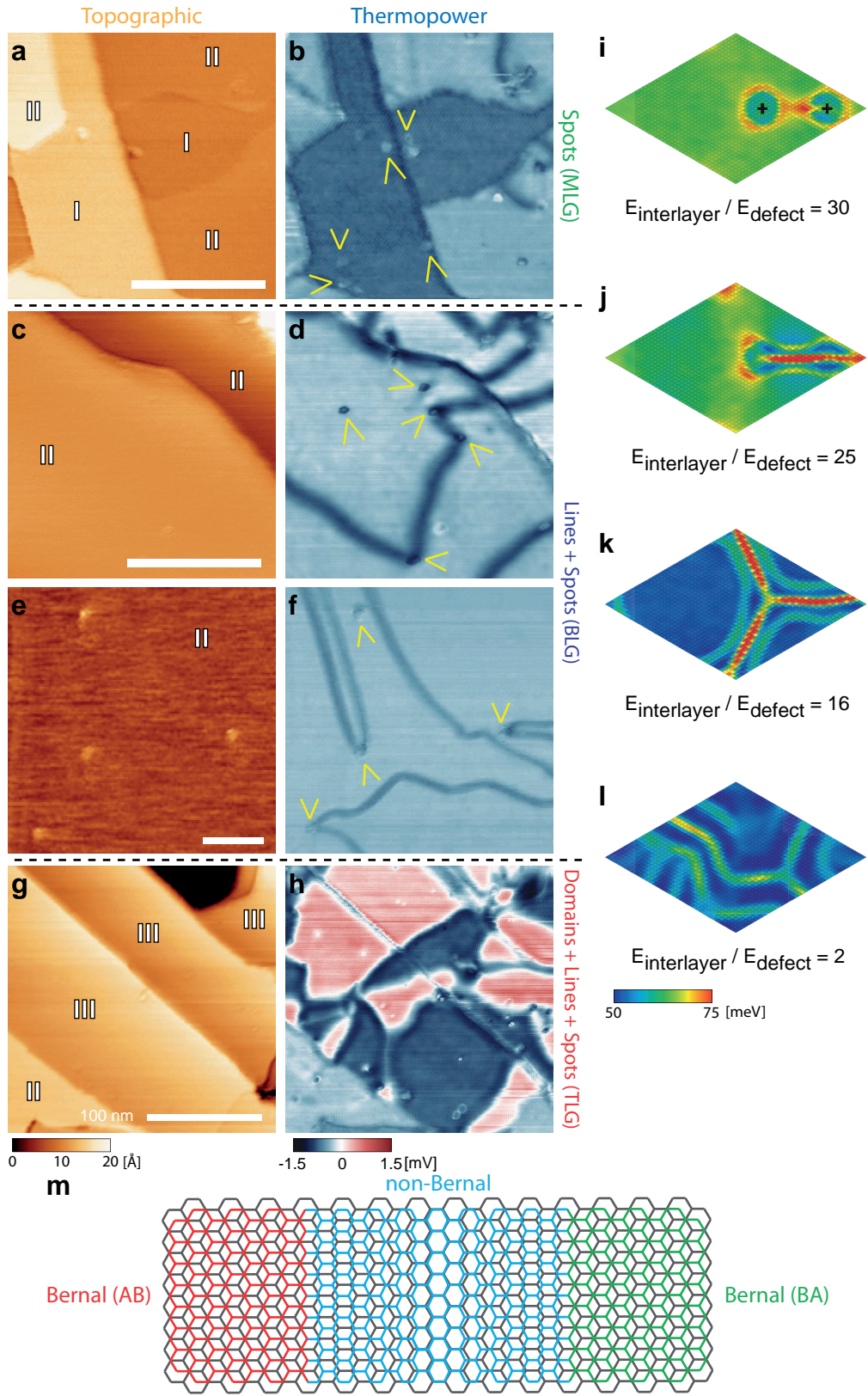

**Figure 2 | Evolution of strain-response patterns in epitaxial graphene. a-h,** Topographic (**a,c,e,g**) and corresponding thermopower (**b,d,f,h**) images. **a, b** are from MLG; **c-f** are from BLG; **g,h** are from TLG. In a flat region of BLG (**e,f**), the dark spots in **f** (yellow arrows) correspond to the

buckled spots in the topographic image, **e**. In TLG, a patched pattern with domains of opposite-signed thermopower is found (**g,h**). The indices in the topographic images denote the number of the graphene layers within the terrace. The temperature difference between the cantilever and the sample was around 30 K. **i-l,** Strain energy maps simulated by the Monte Carlo method showing how graphene sheets with (40 x 40) unit cells accommodate a given compressive strain. The colour bar indicates the local strain energy per carbon atom. The sheet was designed to contain two defect spots — black crosses in **i**. The decreasing interlayer cohesive energy ($E_{interlayer}$) with respect to the defect binding energy ($E_{defect}$) emulates the increasing thickness of graphene above the buffer layer. An exaggerated strain of 2 % was applied for illustrative purposes. **m,** A model illustrating a wall between different Bernal stacking domains. The black layer is the bottom layer; red (AB), green (BA) and sky-blue (domain wall) layers are on the top.

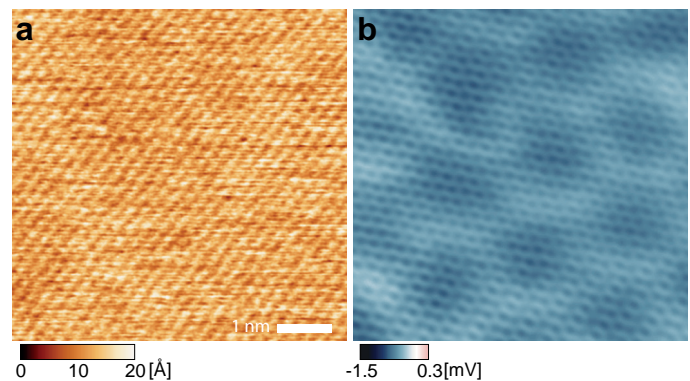

**Figure 3 | Thermoelectric imaging at the atomic scale. a,** AFM topographic image of BLG. **b,** A simultaneously obtained atomically resolved thermoelectric image of BLG acquired by the normal scanning mode. The images were obtained near room temperature with a diamond-coated conducting tip and the temperature difference was approximately 30 K. The interaction between the graphene layer and its substrate is evident from the $6\sqrt{3} \times 6\sqrt{3} R30°$ pattern in **b**.

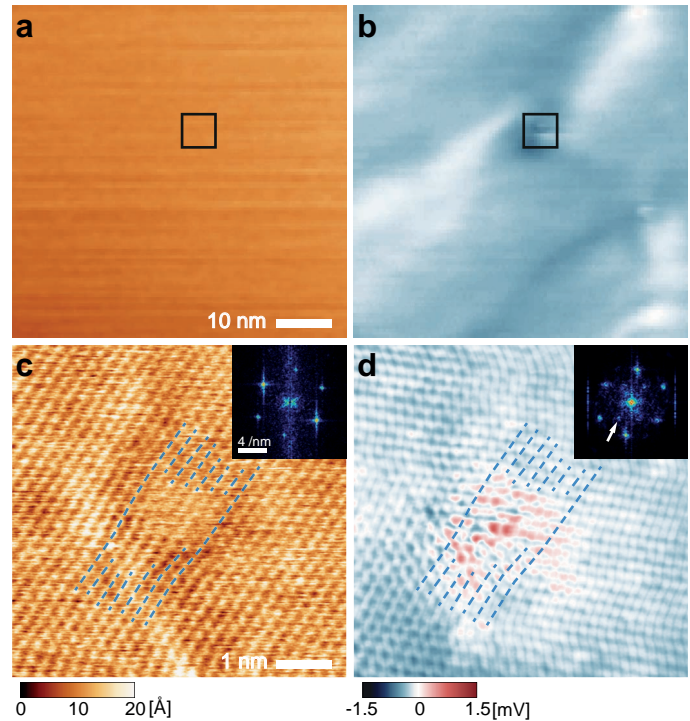

**Figure 4 | Local defect investigation by thermoelectric imaging. a,** AFM topographic image of BLG. **b,** A simultaneously obtained thermoelectric image acquired by the normal scanning mode. **c,** AFM topographic image obtained at the square-region of **a**. The presence of a defect is shown by the change in row numbers, which is indicated by the dotted lines increasing from seven to eight rows across the defect. **d,** The thermoelectric image corresponding to **c**. The insets of **c, d** are the Fourier transform patterns of the images showing the reciprocal lattices. The white arrow in the inset of **d** indicates one of the scattering peaks that appear at the Brillouin zone boundary. The imaging condition was identical to that of Fig. 3, and the base-line of the thermoelectric voltage was calibrated.

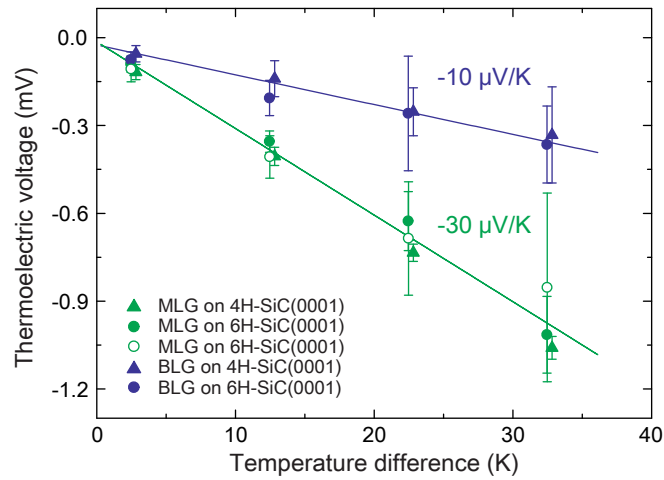

**Figure 5 | Spectroscopic measurements of the thermoelectric voltage from various graphene samples as a function of applied temperature difference.** The data shows linear slopes dependent on the layer thickness but not on growth conditions. Closed and open symbols denote BLG- and MLG-dominant samples, respectively. Data points are an average from a distribution of measurements collected from stationary point measurements. The error bars indicate the full-width at half-maximum of the measurement distributions. The intersection of the linear plots around 0 mV shows that the laser heating on the cantilever has minimal effects. Measured values had little dependency on the force setpoint up to 30 nN. The reservoir temperature ($\Delta T=0$) was at room temperature.

# Supplementary Information

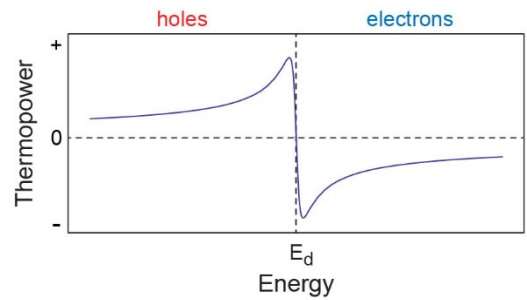

**Figure S1 |** A schematic showing how the thermopower of pristine graphene depends on the position of the Fermi-level.

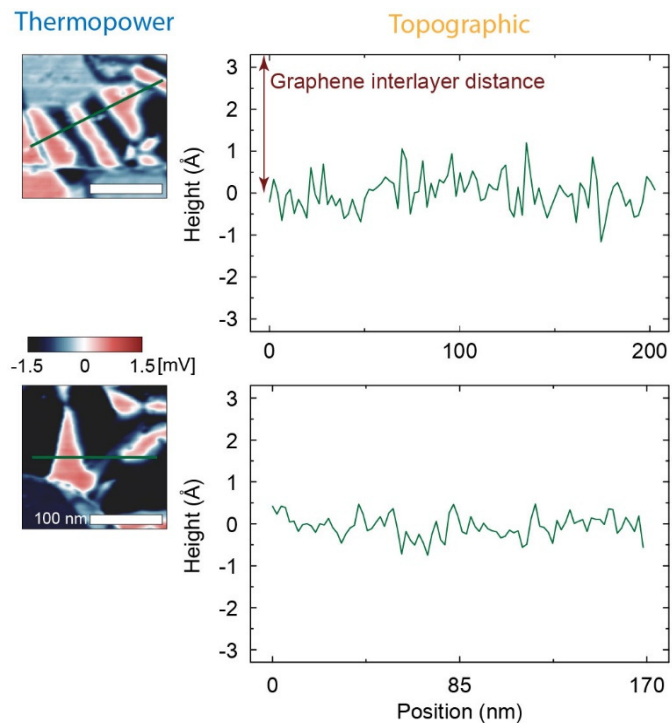

**Figure S2 |** Topographic profiles across the positive- and negative-thermopower domains in TLG. Each profile corresponds to the green line marked on the thermopower image. The fluctuation level in the profile was well below the graphene interlayer distance (≈3.3 Å). The scale bars in the thermopower images both represent a length of 100 nm.

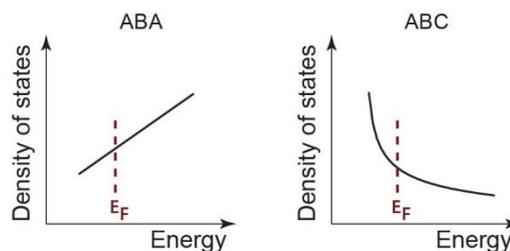

**Figure S3 |** Schematic low-energy DOS profiles of TLG for ABA and ABC stacking. The Fermi-level in TLG has been reported to be ≈0.2 eV above $E_d$ when grown on SiC[34].

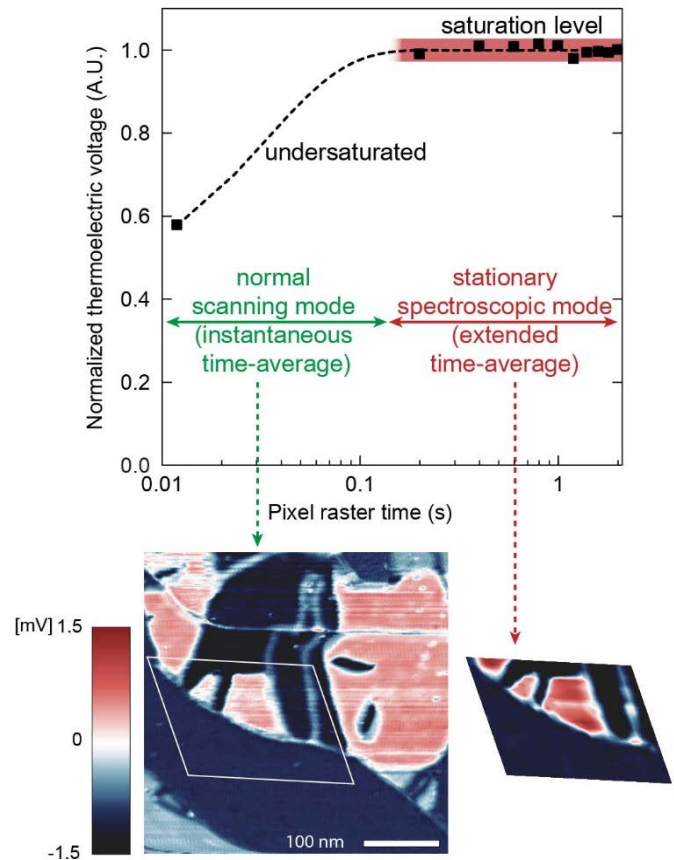

**Figure S4** | Comparison of the normal and stationary spectroscopic modes. The images obtained from the same area are qualitatively similar, while the spatial resolution is better in the normal scan image and the absolute value of the thermoelectric voltage is more reliable in the stationary scan image.

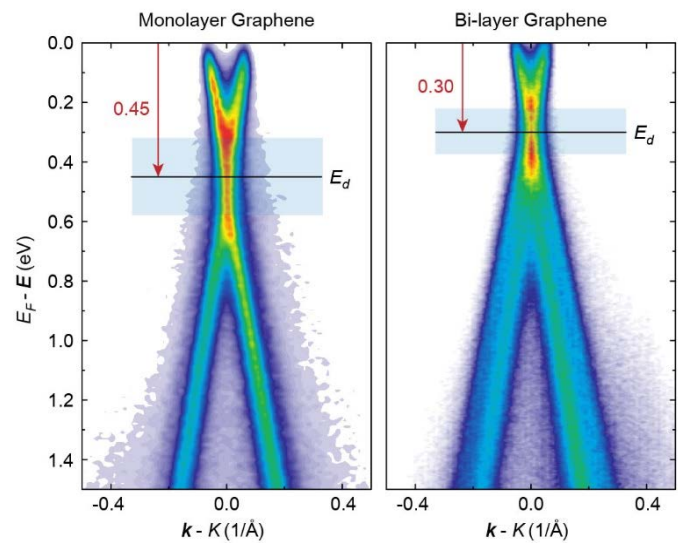

**Figure S5** | ARPES intensity maps obtained at **k** sections across the $K$ point in epitaxial graphene. The Dirac point levels $E_d$ are indicated with respect to the Fermi-levels, and the energy gaps found from **k** section-plots are depicted as blue-shaded boxes around $E_d$. The analysis procedure is identical to that found in ref.[34].

*Simplified Version (Please see http://dx.doi.org/10.1038/nmat3708 for the full content)*

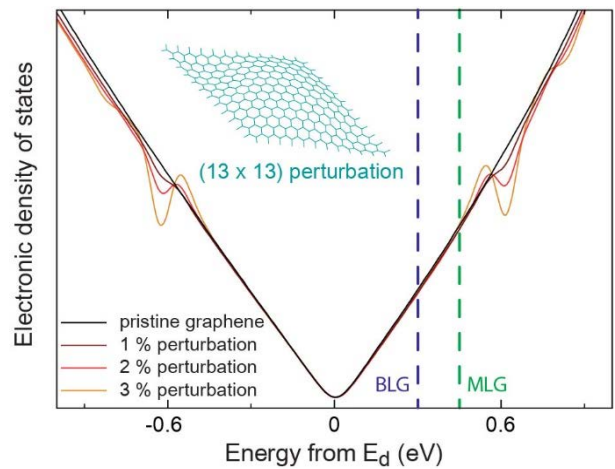

**Figure S6 |** Modification in the electronic DOS of graphene due to the zone boundary scattering in a (13 × 13) supercell. The position of the DOS modification was found at 0.5 eV to 0.6 eV above the Dirac point. A compressive strain of 1 % to 3 % was applied as the perturbation source for the (13 × 13) periodicity. The inset is the perturbed model used for calculation, and the dotted lines indicate the Fermi level positions in our epitaxial MLG and BLG samples.

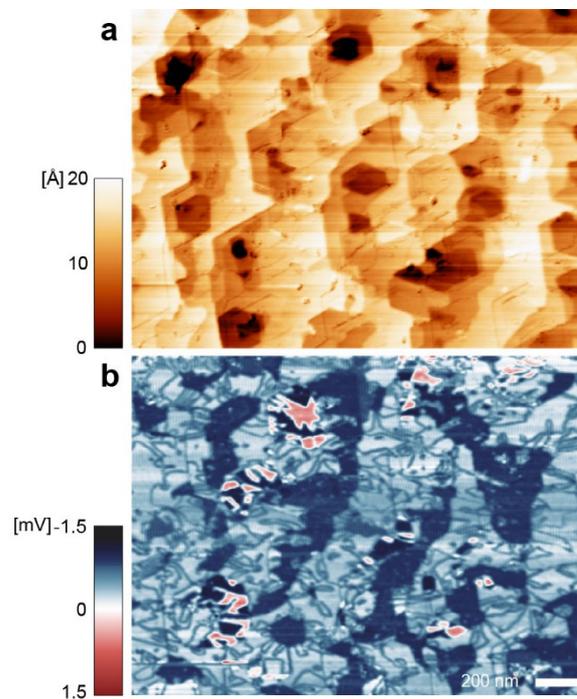

**Figure S7 | a,** Topographic and **b,** thermopower image simultaneously obtained from an area of 2 μm length.

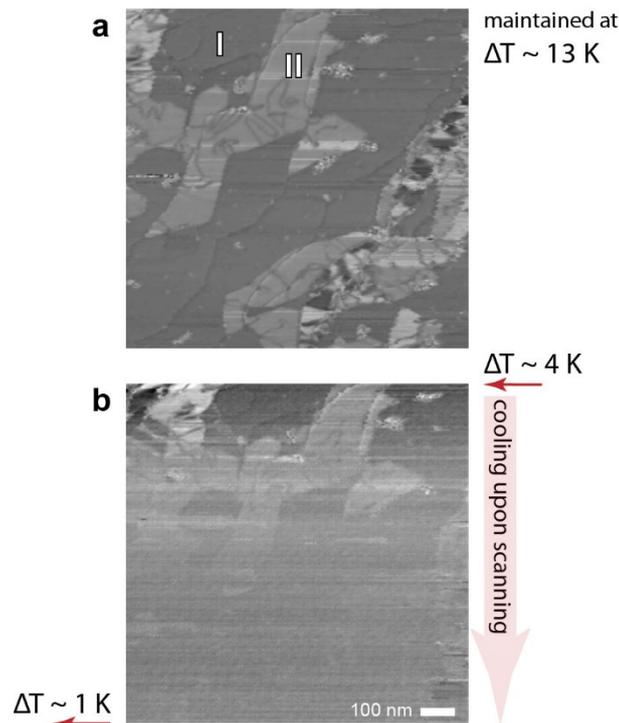

**Figure S8 |** A thermopower image of graphene obtained upon reducing the temperature difference between the sample and the cantilever **(b)**, compared with an image obtained at a maintained temperature difference of ≈13 K **(a)**. In **b** the sample was cooled down from 301 K (scan initiated, right-top side of the image) to 298 K (scan finished, left-bottom side of the image) while the cantilever was maintained at 297 K. The contrast between MLG (region I) and BLG (region II) vanished upon cooling, along with the line patterns in BLG. The grayscale is to show the relative contrast within the image, and thus is arbitrary.

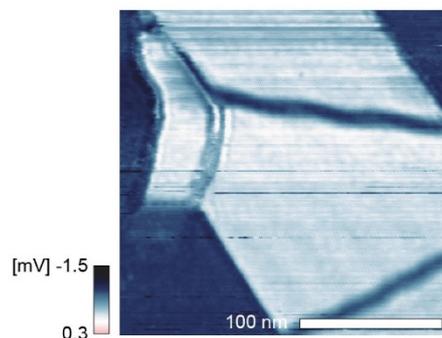

**Figure S10 |** STM images of **a**, MLG and **b**, BLG, obtained at room temperature from the identical sample used in our thermoelectric imaging experiment. The tunnelling current was 100 pA for both images, and the samples were each biased at -100 mV (MLG) and -200 mV (BLG).

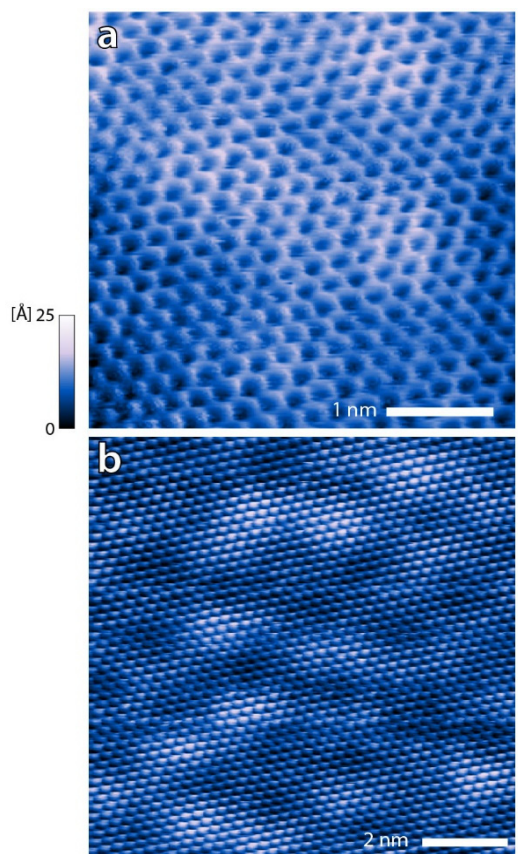

**Figure S10 |** A thermopower image showing a BLG region with line patterns. The sample is a MLG-dominant sample grown on 6H-SiC(0001) by furnace-annealing. The temperature difference applied for imaging was approximately 30 K.